\documentclass[aps,prl,reprint,showpacs,lengthcheck,longbibliography,superscriptaddress]{revtex4-1}
\usepackage[T1]{fontenc}

\pdfoutput=1

\usepackage{graphicx}
\usepackage{float}
\usepackage{amsmath}
\usepackage{amssymb}
\usepackage[usenames,dvipsnames]{xcolor}
\usepackage[colorlinks,linkcolor=blue,citecolor=blue,urlcolor=blue]{hyperref}
\usepackage{microtype}

\def\equationautorefname~#1\null{Equation~(#1)\null}

\begin{document}

\title{Enhancing Photon Correlations through Plasmonic Strong Coupling}
\author{R. S\'aez-Bl\'azquez}
\affiliation{Departamento de F\'isica Te\'orica de la Materia
Condensada and Condensed Matter Physics Center (IFIMAC),
Universidad Aut\'onoma de Madrid, E- 28049 Madrid, Spain}
\author{J. Feist}
\affiliation{Departamento de F\'isica Te\'orica de la Materia
Condensada and Condensed Matter Physics Center (IFIMAC),
Universidad Aut\'onoma de Madrid, E- 28049 Madrid, Spain}
\author{A. I. Fern\'andez-Dom\'inguez}\email{a.fernandez-dominguez@uam.es}
\affiliation{Departamento de F\'isica Te\'orica de la Materia
Condensada and Condensed Matter Physics Center (IFIMAC),
Universidad Aut\'onoma de Madrid, E- 28049 Madrid, Spain}

\author{F. J. Garc\'ia-Vidal}\email{fj.garcia@uam.es}
\affiliation{Departamento de F\'isica Te\'orica de la Materia
Condensada and Condensed Matter Physics Center (IFIMAC),
Universidad Aut\'onoma de Madrid, E- 28049 Madrid, Spain}
\affiliation{Donostia International Physics Center (DIPC), E-20018
Donostia/San Sebasti\'an, Spain}

\begin{abstract}
\vspace{1mm}
There is an increasing scientific and technological interest on
the design and implementation of nanoscale sources of quantum
light. Here, we investigate the quantum statistics of the light
scattered from a plasmonic nanocavity coupled to a mesoscopic
ensemble of emitters under low coherent pumping. We present an
analytical description of the intensity correlations taking place
in these systems, and unveil the fingerprint of
plasmon-exciton-polaritons in them. Our findings reveal that
plasmonic cavities are able to retain and enhance excitonic
nonlinearities even when the number of emitters is large. This
makes plasmonic strong coupling a promising route for generating
nonclassical light beyond the single emitter level.
\vspace{2cm}
\end{abstract}

\maketitle

\section{Introduction}
Much research attention has focused lately on plasmonic
nanocavities for strong coupling applications. In these devices,
the interaction between surface plasmons (SPs) and quantum
emitters (QEs) can be intense enough to yield new hybrid
light-matter states, the so-called plasmon-exciton-polaritons
(PEPs)~\cite{Torma2015}. PEPs involving macroscopic QE ensembles
have been reported in
planar~\cite{Bellesa2004,Schwartz2011,GonzalezTudela2013} and
nanoparticle~\cite{Delga2014,Zengin2015,Todisco2015} geometries,
and they have been used for controlling chemical
reactions~\cite{Hutchison2012,Galego2015} or enhancing
charge/energy transport~\cite{Feist2015,Orgiu2015}. From a purely
photonics perspective, room temperature PEP lasing has been
recently reported~\cite{Hakala2017,Ramezani2017}. However, in
order to harness the full potential of plasmonic cavities for
quantum optical applications, plasmonic systems that display
nonlinearities at the single-photon level would be highly
desirable~\cite{Tame2013}. This is not possible in macroscopic
ensembles, which present collective boson-like behavior at pumping
levels below the QE saturation regime~\cite{Holstein1940}.

Very recently, strong-coupling signatures in the power spectrum of
nano-gap metallic cavities filled with only a few QEs have been
reported~\cite{Chikkaraddy2016,Santhosh2016}. These experimental
advances have been accompanied by theoretical efforts aiming to
clarify the near-field conditions yielding PEPs at the single
emitter level~\cite{Li2016}. However, the generation of
nonclassical light through plasmonic strong coupling has not been
explored yet. In this Article, we fill this gap by investigating
the quantum statistics of the photons scattered by a nanocavity
strongly coupled to a mesoscopic emitter ensemble (up to $\sim100$
QEs) under coherent pumping. We develop an analytical description
of the quantum optical properties of the system that allows us to
reveal that, contrary to what is expected, plasmonic cavities
enhance photon correlations in QE ensembles of considerable size
under strong coupling conditions.

\begin{figure}[h]
	\includegraphics[width=\linewidth]{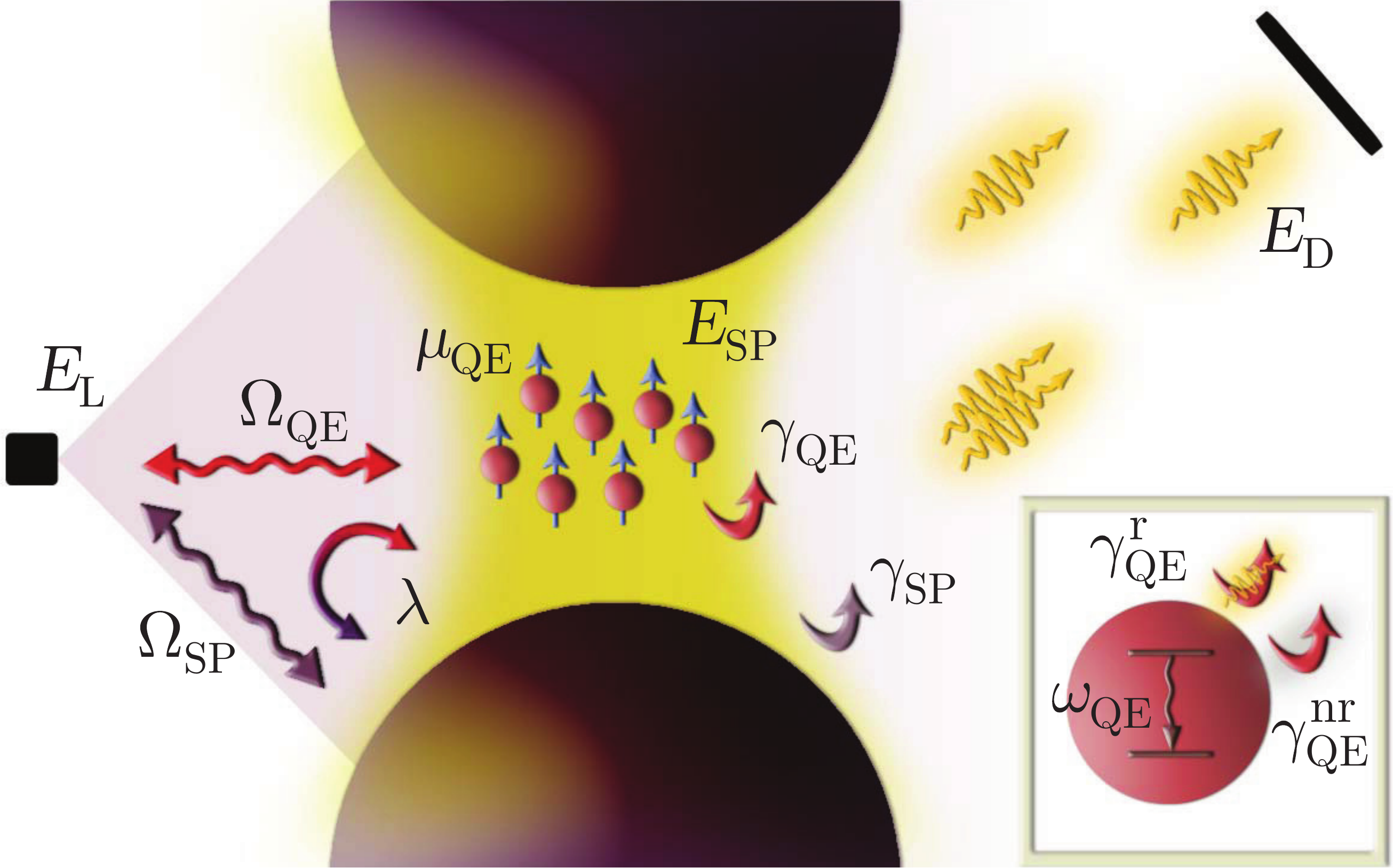}
	\caption{A QE ensemble resonantly coupled to a generic plasmonic
		cavity. The right inset depicts the two-level QE model.}
	\label{fig:1}
\end{figure}

\section{Model}

\autoref{fig:1} depicts the system under study: $N$ identical QEs
with transition dipole moment $\boldsymbol{\mu}_{\rm QE}$ and
frequency $\omega_{\rm QE}$ interact with the near-field ${\bf
	E}_{\rm SP}$ (the same for all QEs) of a single SP mode of energy
$\omega_{\rm SP}$ supported by a generic nanocavity. Both
subsystems undergo radiative and nonradiative damping, with decay
rates $\gamma_{\rm QE/SP}=\gamma_{\rm QE/SP}^{\rm r}+\gamma_{\rm
	QE/SP}^{\rm nr}$. We consider QEs in which pure dephasing is
negligible as this process would suppress quantum correlations in
the emitted photons. Both QEs and SP are coherently driven by a
laser field ${\bf E}_{\rm L}$ with frequency $\omega_{\rm L}$. The
steady-state density matrix $\hat{\rho}$ for the hybrid system is
the solution of the Liouvillian equation ($\hbar=1$)
\begin{equation}
	i [\hat{\rho},\hat{H}]+\frac{\gamma_{\rm
			SP}}{2}\mathcal{L}_{\hat{a}}[\hat{\rho}]+\frac{\gamma_{\rm
			QE}^{\rm
			r}}{2}\mathcal{L}_{\hat{S^-}}[\hat{\rho}]+\frac{\gamma_{\rm
			QE}^{\rm
			nr}}{2}\sum_{i=1}^N\mathcal{L}_{\hat{\sigma}_i}[\hat{\rho}]=0,\label{Liouvillian}
\end{equation}
where $\hat{a}$, $\hat{\sigma}_i$, and
$\hat{S}^-=\sum_{i=1}^N\hat{\sigma}_i$ are the annihilation
operators for the SP mode, the $i$-th QE, and the ensemble
super-radiant state, respectively. The damping associated to
operator $\hat{O}$ is described by standard Lindblad
super-operators $\mathcal{L}_{\hat{O}}[\hat{\rho}] = 2\hat{O}
\hat{\rho} \hat{O}^\dag - \{\hat{O}^\dag \hat{O}, \hat{\rho}\}$.
~\eqref{Liouvillian} reflects that, contrary to nonradiative
decay, radiation damping is a coherent process which involves only
the super-radiant state of the QE ensemble (the rest of the
ensemble states are dark). In the rotating frame, the coherent
dynamics is governed by the time-independent Tavis-Cummings
Hamiltonian~\cite{Tavis1968}
\begin{eqnarray}
	\hat{H}&=& \Delta_{\rm SP} \hat{a}^\dag\hat{a} + \Delta_{\rm QE}
	\hat{S}_z
	+\lambda (\hat{S}^+\hat{a}+\hat{S}^-\hat{a}^\dag)+ \nonumber\\
	&&+\Omega_{\rm SP}(\hat{a}^\dag+\hat{a})+\Omega_{\rm
		QE}(\hat{S}^++\hat{S}^-),\label{Hamiltonian}
\end{eqnarray}
with $\Delta_{\rm QE/SP}=\omega_{\rm QE/SP}-\omega_{\rm L}$ and
$\hat{S}_z=\tfrac{1}{2}[\hat{S}^+,\hat{S}^-]$. The QE-SP coupling
is $\lambda={\bf E}_{\rm SP}\cdot\boldsymbol{\mu}_{\rm QE}$, while
$\Omega_{\rm QE}={\bf E}_{\rm L}\cdot\boldsymbol{\mu}_{\rm QE}$
and $\Omega_{\rm SP}={\bf E}_{\rm L}\cdot\boldsymbol{\mu}_{\rm
	SP}$ are the pumping frequencies. Here, $\boldsymbol{\mu}_{\rm
	SP}$ is the effective SP dipole moment. Once the steady-state
density matrix is known, the first- and second-order correlation
functions can be calculated from the scattered far-field operator
at the detector ${\bf \hat{E}}_{\rm
	D}^-\propto\boldsymbol{\mu}_{\rm SP}\hat{a}^\dag+
\boldsymbol{\mu}_{\rm QE}\hat{S}^+$. Note that we have taken
advantage of the sub-wavelength dimensions of the system to
neglect the differences between the electromagnetic Green's
function describing the emission from the SP and the various QEs
in ${\bf \hat{E}}_{\rm D}^-$.

\section{Results and discussion}

\begin{figure}[!t]
	\includegraphics[width=\linewidth]{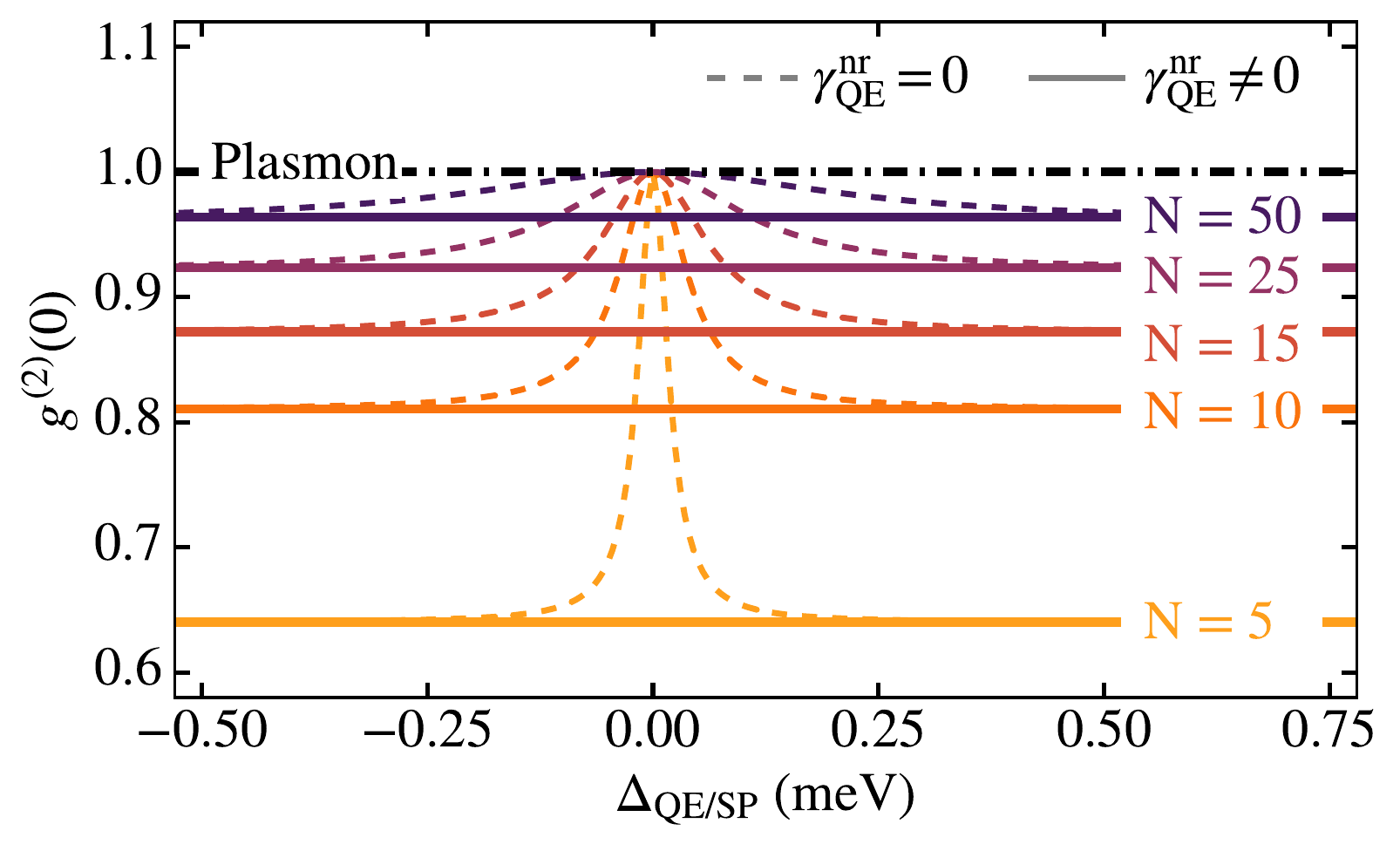}
	\caption{Correlation function $g^{(2)}(0)$ versus laser detuning
		for SP (black dash-dotted line) and QEs (color lines) uncoupled.
		Various ensembles sizes are shown, with (solid) and without
		(dashed) the inclusion of QE nonradiative decay, $\gamma^{\rm
			nr}_{\rm QE}$.} \label{fig:2}
\end{figure}

Before investigating photon correlations under strong coupling
conditions, we consider first both SP and QE subsystems uncoupled.
For this purpose, we solve \autoref{Liouvillian} numerically and
compute the normalized zero-delay second-order correlation
function in the steady state ${g^{(2)}(0)=\langle{{\bf
			\hat{E}}^-_{\rm D}}{\bf \hat{E}}^-_{\rm D}{\bf \hat{E}}_{\rm
		D}^+{\bf \hat{E}}_{\rm D}^+\rangle/\langle{\bf \hat{E}}^-_{\rm
		D}{\bf \hat{E}}^+_{\rm D}\rangle^2}$. This magnitude measures the
intensity fluctuations of the emitted light, and is related to the
probability for two photons to arrive at the same time at the
detector. Values of $g^{(2)}(0)$ smaller than one indicate
antibunching, which cannot be achieved with classical
light~\cite{Scully1997}. We only consider low laser intensities,
and study quantum correlations far from the pumping regime in
which QE saturation becomes relevant. \autoref{fig:2} plots
$g^{(2)}(0)$ as a function of the laser detuning for an empty
plasmonic cavity (black dash-dotted line) and ensembles of
different number of emitters (color solid lines). For comparison,
the correlation spectra for QEs with $\gamma^{\rm nr}_{\rm QE}=0$
are also shown (color dashed lines). The parameters modelling the
single SP mode are: $\omega_{\rm SP}=3$ eV, $\gamma_{\rm
	SP}=0.1$~eV and $\mu_{\rm SP}=19$ e$\cdot$nm~\cite{Delga2014}. Our
calculations yield $g^{(2)}(0)=1$, as expected from the SP
inherent bosonic character. For proof-of-principle purposes, we
have chosen QE parameters as: $\omega_{\rm QE}=3$ eV, $\gamma^{\rm
	r}_{\rm QE}=6$ $\mu$eV ($\mu_{\rm QE}=1$ e$\cdot$nm), and
$\gamma^{\rm nr}_{\rm QE}=15$ meV. These values correspond to
organic molecules that display very low quantum yield and in which
collective strong coupling has been already
reported~\cite{Schwartz2011,Ramezani2017}. As we show below, this
type of QEs are also favorable for generating photon correlations.
Notice then that for a practical realization of our findings with
organic QEs, the experiments should be carried out at low
temperature in order to avoid pure dephasing processes inside the
QEs. For all $N$, photon statistics is sub-Poissonian
($g^{(2)}(0)<1$), but the degree of antibunching decreases rapidly
with the ensemble size. As $N$ increases, the system bosonizes and
the quantum character of the scattered light is lost (note that
$g^{(2)}(0)=0.96$ for $N=50$). Neglecting nonradiative damping
only leads to an extremely narrow Lorentzian-like profile, which
suppresses antibunching exactly at zero detuning. This observation
is in agreement with the resonance fluorescence phenomenology of a
single QE~\cite{Mollow1969}, in which no incoherent scattering
occurs in the limit of vanishing pumping (saturation effects in
the QE population are negligible). Note that the $g^{(2)}(0)$
behaviour obtained from our model is in accordance with more
sophisticated descriptions~\cite{Miftasani2016} of QE ensembles.

\begin{figure*}[!t]
	\includegraphics[width=\textwidth]{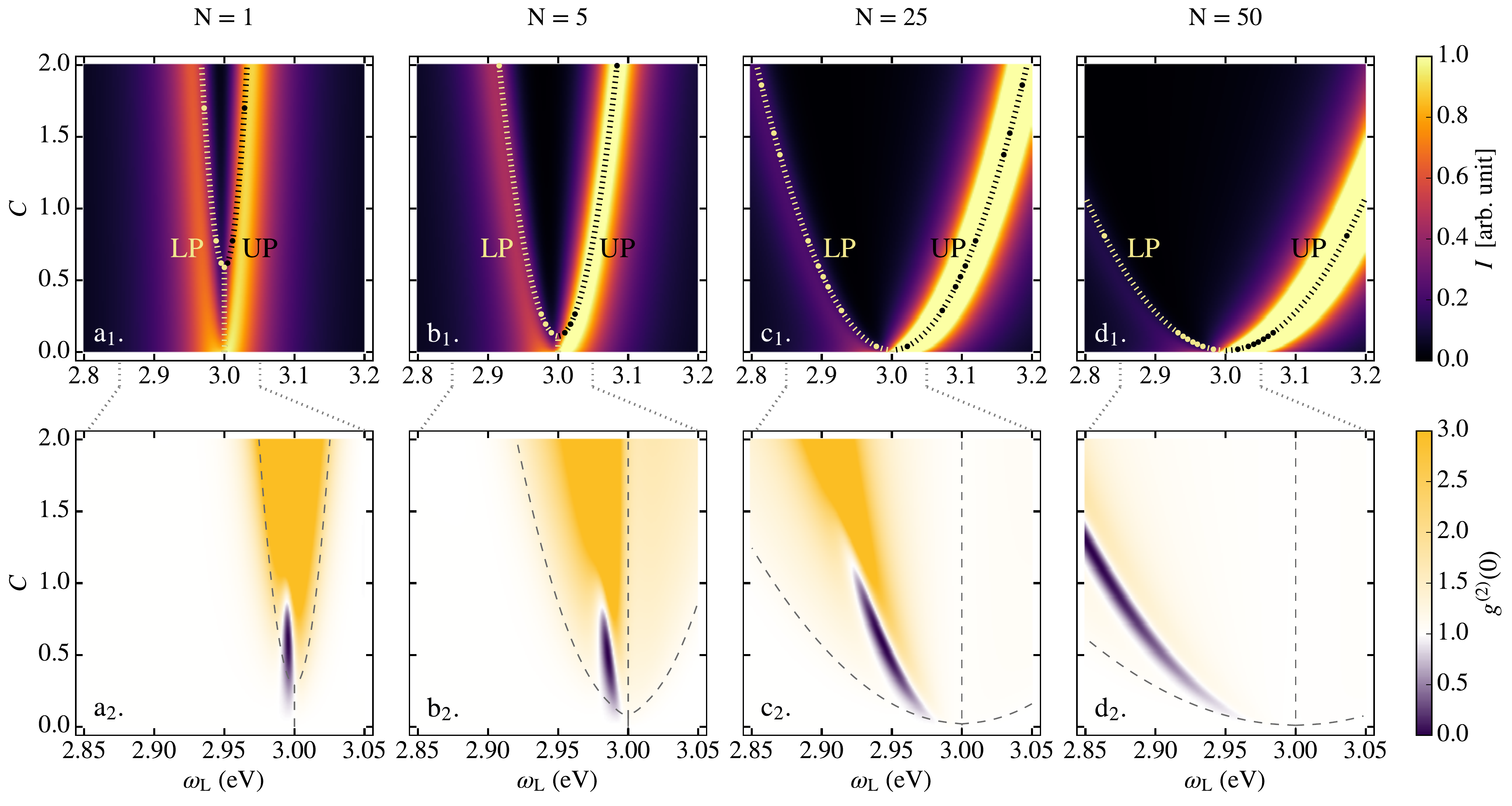}
	\caption{Scattering intensity $I$ (${\rm a_{1}}$-${\rm d_{1}}$)
		and correlation function $g^{(2)}(0)$ (${\rm a_{2}}$-${\rm
			d_{2}}$) versus laser frequency and single emitter cooperativity
		for various QE-SP systems. In the upper (lower) panels dotted
		(dashed) lines plot the PEP frequencies (half frequencies) in the
		one-excitation (two-excitation) manifold.} \label{fig:3}
\end{figure*}

Exact numerical solutions to \autoref{Liouvillian} can be obtained
for strong QE-SP coupling. However, such calculations are only
possible for configurations involving very small QE
ensembles~\cite{Auffeves2011}, even far from the QE saturation
regime~\cite{Poddubny2010}. In order to circumvent this limitation
and explore photon statistics in mesoscopic ensembles, we map
\autoref{Liouvillian} into the effective non-Hermitian
Hamiltonian~\cite{Visser1995}
\begin{eqnarray}
	\hat{H}_{\rm eff}&=&\hat{H}-i\frac{\gamma_{\rm SP}}{2}
	\hat{a}^\dag\hat{a} -i\frac{\gamma^{\rm nr}_{\rm QE}}{2} \hat{S}_z
	-i\frac{\gamma^{\rm r}_{\rm QE}}{2}
	\hat{S}^+\hat{S}^-,\label{Effective}
\end{eqnarray}
where $\hat{H}$ is given by \autoref{Hamiltonian}. Note that
$\hat{H}_{\rm eff}$ depends only on the collective bright state
operators of the QEs and is independent of the dark states of the
ensemble (superpositions of QE excitations that do not couple to
the plasmon or external light), which means a drastic reduction in
the Hilbert space for large $N$. \autoref{Effective} results from
neglecting the refilling terms $\hat{O} \hat{\rho} \hat{O}^\dag$
in the Lindblad super-operators in \autoref{Liouvillian}. This
approach can be safely employed in the regime of low pumping,
where the ground state can be considered as a reservoir with
population equal to $1$. In this limit, we can solve the
Schr\"odinger equation for $\hat{H}_{\rm eff}$ treating the
coherent driving, $E_{\rm L}$, as a perturbative
parameter~\cite{Brecha1999}. More details on the effective
Hamiltonian approach can be found in Supplemental Material.

As we are interested in intensity correlations, we can restrict
our perturbative treatment of \autoref{Effective} to second order
and truncate the Hilbert space at two excitations.
In the following, for
simplicity, we also assume the plasmonic near-field, ${\bf E}_{\rm
	SP}$, parallel to the laser field, ${\bf E}_{\rm L}$ (as, for
example, in particle-on-mirror cavities~\cite{Chikkaraddy2016}).
Moreover, we only consider the optimum configuration for strong
coupling, in which $\boldsymbol{\mu}_{\rm QE}$ is aligned with
${\bf E}_{\rm SP}$. The scattering intensity, $I=\langle{\bf
	\hat{E}}^-_{\rm D}{\bf \hat{E}}^+_{\rm D}\rangle$, is given within
first-order perturbation theory as
\begin{equation}
	I=(\eta N \mu_{\rm SP}\Omega_{\rm
		SP})^2\left|\frac{\eta\tilde{\Delta}_{\rm SP}+\tilde{\Delta}_{\rm
			QE}/\eta N-2\lambda}{\tilde{\Delta}_{\rm SP}\tilde{\Delta}_{\rm
			QE}-N\lambda^2}\right|^2, \label{Intensity}
\end{equation}
where $\eta=\mu_{\rm QE}/\mu_{\rm SP}=\Omega_{\rm QE}/\Omega_{\rm
	SP}$, $\tilde{\Delta}_{\rm SP}=\Delta_{\rm SP}-i\gamma_{\rm SP}/2$
and $\tilde{\Delta}_{\rm QE}=\Delta_{\rm QE}-i(\gamma^{\rm
	nr}_{\rm QE}+N\gamma^{\rm r}_{\rm QE})/2$. Using second-order
perturbation theory, the correlation function, $g^{(2)}(0)$, can
be expressed as \vspace{-21mm}
\section*{{}}
\begin{widetext}
	\begin{equation}
		g^{(2)}(0)=\left|1-\frac{1}{N} \left(\frac{\eta\tilde{\Delta}_{\rm
				SP}-\lambda}{\eta\tilde{\Delta}_{\rm SP}+ \tilde{\Delta}_{\rm
				QE}/\eta N-2\lambda}\right)^2\frac{(\tilde{\Delta}_{\rm
				QE}+iN\gamma^{\rm r}_{\rm QE}/2)[\tilde{\Delta}_{\rm
				QE}\tilde{\Delta}_{\rm SP}+(\tilde{\Delta}_{\rm
				SP}-\lambda/\eta)^2-N\lambda^2]}{(\tilde{\Delta}_{\rm
				QE}+i\gamma^{\rm r}_{\rm QE}/2)[\tilde{\Delta}_{\rm
				QE}\tilde{\Delta}_{\rm SP}+\tilde{\Delta}_{\rm
				SP}^2-N\lambda^2]-\tilde{\Delta}_{\rm SP}(N-1)\lambda^2}\right|^2.
		\label{g2completa}
	\end{equation}
\end{widetext}
Note that, for $\gamma_{\rm QE}^{\rm nr}\gg\gamma_{\rm QE}^{\rm
	r}$, \autoref{g2completa} yields $g^{(2)}(0)=(1-1/N)^2$ at
$\lambda=0$ and $\eta\rightarrow\infty$, recovering the flat
correlation spectra in \autoref{fig:2} for low-quantum-yield QE
ensembles.

\autoref{fig:3} renders the far-field intensity (top row) and
correlations (bottom row) for a nanocavity filled with four
different QE ensembles: $N=1$ (a), 5 (b), 25 (c) and 50 (d). The
horizontal and vertical axes correspond to laser frequency and
QE-SP coupling strength, respectively. The latter is expressed
through the single-emitter cooperativity,
$C=2\lambda^2/\gamma_{\rm QE}\gamma_{\rm SP}$, with upper limit
$C=2$ ($\lambda=0.03$ eV), well below the collective ultra-strong
coupling regime. We restrict our attention to QE-SP resonant
coupling and consider the same parameters as in \autoref{fig:2}.
Although the quantitative results shown in \autoref{fig:3} depend
on the specifics of the system, we have checked that our findings
and their fundamental implications remain valid for a wide range
of realistic configurations. As shown in Supplemental Material, the
behavior is also very similar when the SP field is spatially
inhomogeneous or when inhomogeneous broadening is introduced for
the QEs (note that the emitters cannot be formally described
through a single bright state in these cases, but must be treated
individually). Dipole-dipole interactions among QEs are also
analyzed in Supplemental Material. Our results reveal that these have a
significant impact on photon correlations in dense QE ensembles.
Interestingly, we find that antibunching is more robust than
bunching when interactions become large.

The complex $g^{(2)}(0)$ patterns in \autoref{fig:3}(${\rm
	a}_2$)-(${\rm d}_2$) reveal that both photon bunching and
antibunching occur in the strong coupling regime. These panels
also show that the main quantum statistical features emerging at
the single-emitter level (which are in qualitative agreement with
recent experimental reports on semiconductor
cavities~\cite{Faraon2008,Muller2015}) are mostly retained as $N$
increases. Up to $N\sim25$, photon emission is antibunched within
a narrow frequency window located at $C\lesssim1$, which implies
that the single-emitter cooperativity can be considered as the key
parameter determining photon correlations in ensembles containing
up to several tens of QEs~\cite{Habibian2011}. Notice also that,
as a difference with high-quantum-yield QEs in low-loss
semiconductor cavities, the inherent nonradiative losses of
organic molecules and plasmonic systems allow observing
antibunching for large $C$-values (see Supplemental Material for more
details). On the other hand, bunched emission takes place at
larger coupling strengths and within broader spectral domains for
all $N$. Remarkably, there are spectral windows in which strong
antibunching ($g^{(2)}(0) \approx 0$) takes place even for $N=50$,
whereas the emission from the uncoupled QE ensemble is essentially
classical (see \autoref{fig:2}). This is the main result of this
Article, namely that in comparison to the uncoupled subsystems,
collective plasmonic strong coupling can significantly enhance
photon correlations in mesoscopic PEP systems.

By taking advantage of our analytical approach, we can gain
physical insight into the results shown in \autoref{fig:3}. The
intensity maps present two scattering maxima, whose origin lies at
the denominator of \autoref{Intensity}. Its vanishing condition
yields analytical expressions for the dispersion of the lower (LP)
and upper (UP) PEPs in the first rung of the Tavis-Cummings
ladder. These PEP frequencies, which naturally incorporate the
$\sqrt{N}$ scaling characteristic of collective strong coupling,
are plotted in dotted lines in top panels. Note that the intensity
maxima overlap with the PEP dispersion bands except for $N=1$ and
$C\lesssim 0.5$. This region, also perceptible for $N=5$ at lower
$C$, falls within the weak coupling regime, where Fano-like
interferences between SP and QE emission gives rise to sharp
scattering dips~\cite{Ridolfo2010}. As $N$ increases, the contrast
between UP (brighter) and LP (darker) scattering peaks increases.
By introducing the PEP frequencies in the numerator of
\autoref{Intensity}, the origin of this asymmetry becomes clear.
Neglecting QE and SP damping, we obtain
$I\propto(1\mp\sqrt{N}\eta)^2$, where the upper (lower) sign must
be used for LP (UP). Thus, QE and SP dipole moments are
antiparallel along the LP dispersion, which diminishes $I$ as $N$
approaches $1/\eta^2$.

In a similar way as in the scattered intensity, we can expect that
the vanishing of the denominator in the second term of
\autoref{g2completa} could give rise to nonclassical light. At
$N=1$, the resonant frequencies emerging from this condition are
equal to half the energies of the LP (upper sign) and UP (lower
sign) in the second rung of the Jaynes-Cummings
ladder~\cite{Laussy2012}. For $N>1$, the same condition leads to a
cubic equation: it accounts for the emergence of the middle PEPs
in the two-excitation manifold (whose real half-frequency is equal
to $\omega_{\rm QE/SP}$). Moreover, notice the presence of the
numerator of \autoref{Intensity} in the denominator of the first
factor in \autoref{g2completa}. As discussed above, this term
acquires the form $(1-\sqrt{N}\eta)$ at the LP band. Therefore,
the darker character of LPs also makes them more suitable for
photon correlations. PEP half-frequencies in the two-excitation
manifold are plotted in dashed lines in \autoref{fig:3}(${\rm
	a}_2$)-(${\rm d}_2$). The regions of strong photon correlations do
not occur exactly at one of the polariton energies, but slightly
above the LP dispersion. This indicates that photon correlations,
i.e., significant deviations from $g^{(2)}(0)=1$, do not originate
from transitions along a single PEP ladder, but from the
interference in the emission involving different hybrid states.
This underlines the crucial role that strong coupling plays: while
each PEP by itself is quasi-bosonic, the hybridization achieved
through strong coupling ensures the coexistence of multiple mixed
light-matter states separated by the Rabi splitting. It is the
interference between the emission from these different but closely
related states that leads to strongly nonclassical light emission.

\begin{figure}[!t]
	\includegraphics[angle=0,width=\linewidth]{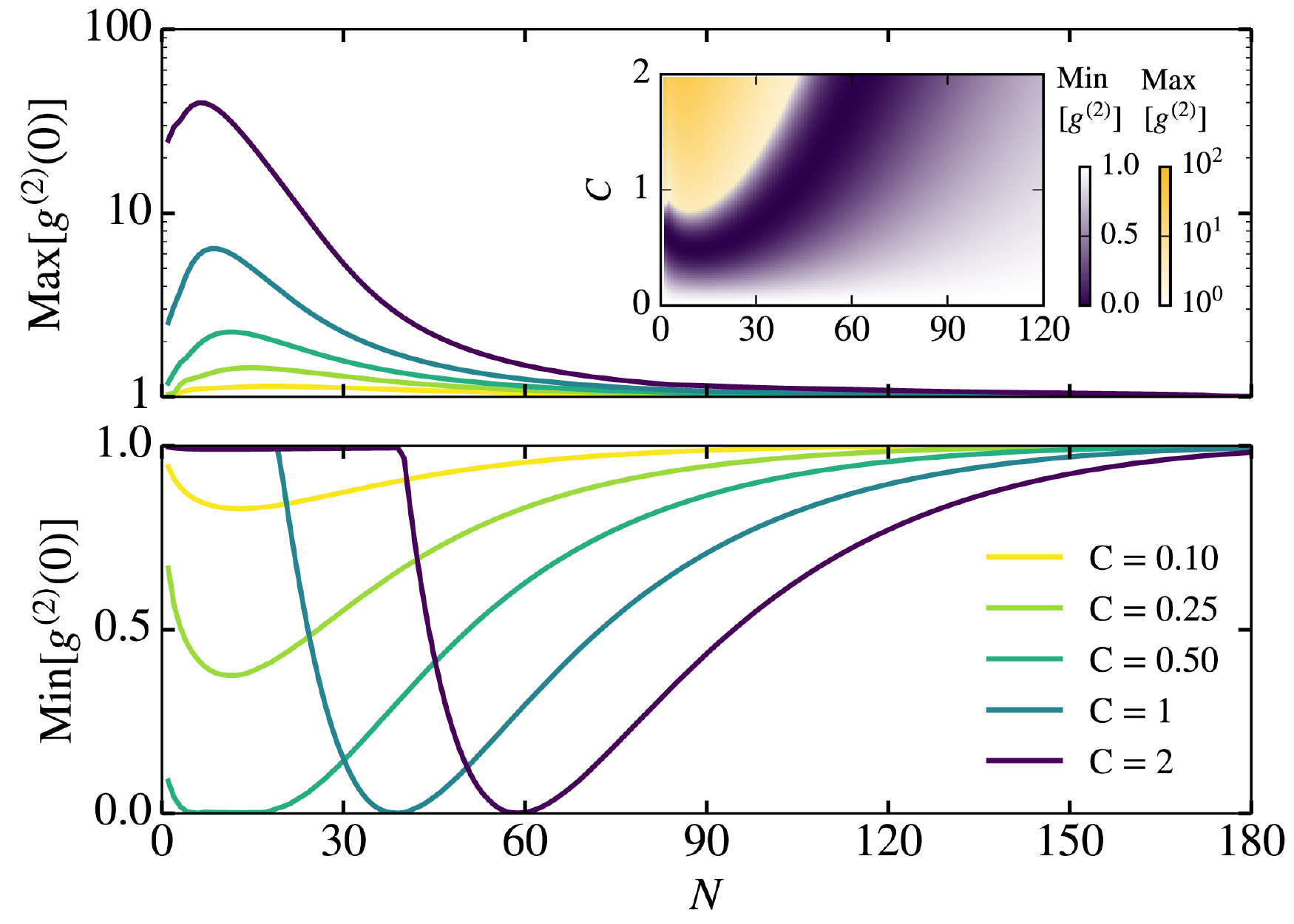}
	\caption{Maximum (top) and minimum (bottom) correlation function
		as a function of the QE ensemble size for several values of the
		single emitter cooperativity. The inset in the upper panel shows
		the map of photon positive (yellow) and negative (violet)
		correlations as a function of $N$ and $C$.} \label{fig:4}
\end{figure}

In order to obtain a general view on the degree of bunching and
antibunching attainable through QE-SP coupling, we evaluate
\autoref{g2completa} at its spectral maxima and minima.
\autoref{fig:4} explore these extreme $g^{(2)}(0)$ values as a
function of cooperativity and number of emitters. The inset
renders overlapping maps for ${\rm Max}[g^{(2)}(0)]$ (yellow) and
${\rm Min}[g^{(2)}(0)]$ (violet), and the top and bottom panels
plot cuts of these maps for various $C$-values. We can identify
three domains according to the statistics of the scattered
photons. For small QE ensembles and large $C$, only positive
correlations take place, as in \autoref{fig:3}(${\rm a}_2$)-(${\rm
	c}_2$) for $C>1$. In this regime, ${\rm Max}[g^{(2)}(0)]$ grows
with increasing coupling strength and develops a maximum at $N\sim
10$ for all $C$. For very large $N$, a second domain is apparent.
In this limit, PEPs bosonize as the $1/N$ factor in
\autoref{g2completa} governs $g^{(2)}(0)$, yielding maxima and
minima approaching $1$ monotonically as the number of QEs
increases. Both bunched and antibunched emission takes place
(within different spectral windows) at intermediate $N$ and $C$.
In this third domain, positive correlations decay monotonically
with $N$, whereas negative correlations are enhanced. ${\rm
	Min}[g^{(2)}(0)]$ diminishes and reaches a minimum value, which
corresponds to the lowest $g^{(2)}(0)$ achievable for a given $N$
and any $C$ (or \textit{vice versa}). It can be proven that this
minimum coincides with a sharp dip in the population of the
plasmon state (written as a linear combination of PEPs) in the
two-excitation manifold. In the limit of vanishing $\eta$ (which
is a good approximation for our problem at small $N$), this
condition simplifies to $C=\tfrac{\gamma_{\rm QE}+\gamma_{\rm
		SP}}{2\gamma_{\rm SP}}\simeq \tfrac{1}{2}$.
\autoref{fig:4}(bottom) shows this minimum developing with
increasing cooperativity at $N\sim 10$ and reaching $g^{(2)}(0)=0$
at $C=0.5$. Remarkably, this zero in $g^{(2)}(0)$ shifts to larger
$N$ for higher cooperativity, yielding strong photon antibunching
at ensemble sizes as large as $100$ QEs. Therefore, as anticipated
in \autoref{fig:3}(${\rm d}_2$), plasmonic strong coupling leads
to the emergence of quantum nonlinearities in large excitonic
systems, which would present $g^{(2)}(0)\simeq1$ when not coupled
to the plasmonic nanocavity.

\vspace{1cm}

\section{Conclusion}

We have investigated the complex photon statistics phenomenology
that emerges from the strong coupling of a mesoscopic ensemble of
quantum emitters and a single plasmon mode supported by a generic
nanocavity. We have presented an analytical method describing the
optical response of these systems under low-intensity coherent
illumination. Our approach provides insights into the role that
both the plasmon-exciton-polariton ladder and its tuning through
the single emitter cooperativity play in the emission of strongly
correlated (bunched and/or antibunched) light. Finally, our
results demonstrate the robustness of these compound systems
against bosonization effects, predicting strong intensity
correlations at considerable ensemble sizes. Our theoretical
findings demonstrate the feasibility and establish experimental
guidelines towards the realisation of nanoscale nonclassical light
sources operating beyond the single-emitter level.

\section{Funding}

This work has been funded by the European Research Council under
Grant Agreement ERC-2011-AdG 290981, the EU Seventh Framework
Programme (FP7-PEOPLE-2013-CIG-630996 and
FP7-PEOPLE-2013-CIG-618229), and the Spanish MINECO under
contracts MAT2014-53432-C5-5-R and FIS2015-64951-R, as well as
through the ``Mar\'ia de Maeztu'' programme for Units of
Excellence in R\&D (MDM-2014-0377).\\

\end{document}